\documentclass[epj]{webofc}
\usepackage[utf8]{inputenc}
\usepackage[varg]{txfonts}   
\usepackage{booktabs}
\usepackage{xcolor}
\definecolor{darkred}{rgb}{0.4,0.0,0.0}
\definecolor{darkgreen}{rgb}{0.0,0.4,0.0}
\definecolor{darkblue}{rgb}{0.0,0.0,0.4}
\usepackage[bookmarks,linktocpage,colorlinks,
    linkcolor = darkred,
    urlcolor  = darkblue,
    citecolor = darkgreen]{hyperref}
\usepackage{subfigure}
\usepackage{bbold}
\usepackage{slashed}
\graphicspath{{./figs/}}
\wocname{EPJ Web of Conferences}
\woctitle{Lattice2017}
\begin{document}
%
\selectlanguage{english}
\title{%
Supermultiplets in \ensuremath{\mathcal{N}}=\ensuremath{1} SUSY SU(2)
   Yang-Mills Theory
}

\author{%
\firstname{Sajid} \lastname{Ali}\inst{1} \and
\firstname{Georg} \lastname{Bergner}\inst{2} \and
\firstname{Henning}  \lastname{Gerber}\inst{1} \fnsep\thanks{Speaker, E-mail:
h.gerber@uni-muenster.de} \and
\firstname{Pietro}  \lastname{Giudice}\inst{1} \and
\firstname{Simon} \lastname{Kuberski}\inst{1} \and
\firstname{Istv{\'a}n}  \lastname{Montvay}\inst{3} \and
\firstname{Gernot}  \lastname{Münster}\inst{1} \and
\firstname{Stefano}  \lastname{Piemonte}\inst{4} \and
\firstname{Philipp}  \lastname{Scior}\inst{1}
}

\institute{%
Institut für Theoretische Physik, Universit\"at Münster,
Wilhelm-Klemm-Str.~9, D-48149 Münster
\and
Theoretisch-Physikalisches Institut, Universit\"at Jena, Max-Wien-Platz
1, D-07743 Jena
\and
Deutsches Elektronen-Synchrotron DESY, Notkestr.~85, D-22603 Hamburg
\and
Fakult\"at für Physik, Universit\"at Regensburg,
Universit\"atsstr.~31, D-93053 Regensburg
}
\abstract{%
We study $\mathcal{N}=1$ supersymmetric Yang-Mills theory (SYM) on
the lattice. The non-perturbative nature of supersymmetric field theories
is still largely unknown. Similarly to QCD, SYM is confining and contains
strongly bound states. Applying the variational method together with different smearing techniques
we extract masses of the lightest bound states such as gluino-glue,
glueball and mesonic states. As these states should form supermultiplets,
this study allows to check whether SYM remains supersymmetric also
on the quantum level. 
}
\maketitle
\section{Introduction}
The Standard Model of particle physics  accurately describes current particle physics experiments. 
However, it is neither theoretically plausible as a fundamental theory (e.g. hierarchy problem), 
nor does it explain the observation of dark matter. It should therefore be viewed as a low energy effective theory of some more complete theory of physics. 
Deviations from the Standard Model are expected to be observed when measuring with more precision  or at higher energies.

One possible extension of the Standard Model which cures the hierarchy problem and also provides candidates for dark matter is supersymmetry. 
Supersymmetry relates bosonic and fermionic fields to one another and implies that for every particle there exists a supersymmetric 
partner particle of the same mass and a spin different by $1/2$. Since this is not observed by current experiments, supersymmetry must be broken if it exists. 

Supersymmetric models have been analyzed extensively within the framework of perturbation theory. On the other hand, it is desirable to gain more insight in their non-perturbative properties.
We study $\mathcal{N}=1$ supersymmetric Yang-Mills theory (SYM) on the lattice. It is the simplest supersymmetric theory containing non-abelian gauge fields.
Important topics to be studied are, besides others, the question whether a supersymmetric continuum limit exists and what the particle spectrum of such a theory looks like.

When treated within the framework of lattice field theory, supersymmetry is broken due to the breaking of Poincar\'{e} invariance by the lattice. 
In Ref. \cite{Curci:1986sm} it has been suggested that  $\mathcal{N}=1$ SYM can be studied on the lattice by extrapolating to the chiral and continuum limit, see \cite{Bergner:2016sbv} for a detailed review.
An important check of supersymmetry restoration in this limit is to study the multiplet formation which was predicted  in Ref. \cite{Veneziano:1982ah} and Ref. \cite{Farrar:1997fn}.
The authors argue that if SYM is unbroken, chiral supermultiplets are formed from glueball states with quantum numbers  $0^{++}$  and
$0^{-+}$ (in $\text{J}^{PC}$ notation), a-$f_0$ and a-$\eta'$ mesons\footnote{The names stem from the similarity of these states with the corresponding particles in QCD.
The fermionic constituents are, however, in the adjoint and not in the fundamental representation of the group, see section \ref{sub:Theory}.} and additional gluino-glue states. 

The groundstates of these particles have been investigated in Ref. \cite{Bergner:2015adz}
for SYM with the gauge group $\mathrm{SU}(2)$. Indeed, a mass degeneracy of these states (except for the $0^{-+}$-glueball) was found in the chiral and continuum limit.
The goal of this work is to analyze also the masses of the corresponding first excited states.

\subsection{The Theory} \label{sub:Theory}
The Lagrangian of SYM in the continuum reads
\begin{equation}
 \mathcal{L}=-\frac{1}{4}F_{\mu\nu}^{a}F^{\mu\nu,a}+\frac{i}{2}\bar{\lambda}^a\left(\slashed{D}\lambda\right)^a-
 \frac{m_{\text{g}}}{2}\bar{\lambda}^a\lambda^a,
\end{equation}
where $F^{\mu\nu,a}$ is the field strength tensor built from the gluon field $A_\mu^a$. The corresponding superpartner is the gluino field $\lambda^{\alpha,a}$. 
Supersymmetry requires the gluino to be in the adjoint representation of the gauge group
and to be of Majorana type, i.e. fulfill the Majorana condition
\begin{equation}
 \bar{\lambda}=\lambda^T C,
\end{equation}
where $C$ is the charge conjugation matrix. The covariant derivative for the gauge group $\mathrm{SU}(2)$ is given by 
\begin{equation}
 \left(D_\mu\lambda\right)^a= \partial_\mu \lambda^a + g \epsilon_{abc} A^b_\mu\lambda^c.
\end{equation}

The gluino mass term breaks supersymmetry softly. In order to approach the chiral limit, the bare gluino mass $m_{\text{g}}$ needs to be fine-tuned such that the renormalized gluino mass vanishes, see also \cite{thiscontrib102}.

For the lattice discretization we use Wilson fermions and a tree level Symanzik improved gauge action. 
Additionally, we use one level of stout smearing \cite{Morningstar:2003gk} for the links in the fermionic part of the action  in order to improve the signal-to-noise ratio of the measurements.

\section{Techniques}

\subsection{Variational Method\label{sub:Variational-Method}}

For the extraction of the excited state masses we use the variational
method as proposed in Ref. \cite{Luscher:1990ck}. It requires solving a generalized eigenvalue problem (GEVP) of the form
\begin{equation}
C(t)\vec{v}^{\ (n)}=\omega^{(n)}(t,t_{0})C(t_{0})\vec{v}^{\ (n)}. \label{eq:GEVP}
\end{equation}
Here $C(t)$ is the correlation matrix, $\omega^{(n)}(t,t_{0})$ and $\vec{v}^{\ (n)}$ are the corresponding generalized eigenvalues and eigenvectors.
The entries of the correlation matrix are given by correlation functions of different interpolating fields $O_i$ of the respective state:
\begin{equation}
C_{ij}( t)=\left\langle O_{i}( t)O_{j}^{\dagger}(0)\right\rangle. \label{eq:Correlation-Matrix}
\end{equation}
The large-$t$ behavior of the generalized eigenvalues allows to extract the mass $m_n$ of the excited states by fitting the eigenvalues $\omega^{(n)}(t,t_0)$ to exponentials:
\begin{equation}
\underset{t\rightarrow\infty}{\lim}\ \omega^{(n)}(t,t_{0})\propto\mathrm{e}^{-m_{n}\left(t-t_{0}\right)}\left(1+\mathcal{O}\left(\mathrm{e}^{-\Delta m_{n}(t-t_{0})}\right)\right),
\quad \Delta m_{n}=\underset{l\ne n}{\min}\left|m_{l}-m_{n}\right|.
\end{equation}
For a good signal it is preferable to choose a set of interpolating fields which on the one hand have good overlaps with the physical states and on the other hand have only small mutual overlaps.

\subsection{Interpolating fields}
The interpolating fields $O_i$ used in the correlation matrix (\ref{eq:Correlation-Matrix}) have to match the quantum numbers of the respective state to be analyzed. 
The basic interpolating fields that we use are described below. In order to improve the signal-to-noise ratio and to create different
interpolating fields for each state we apply different smearing methods to these operators (see Sect. \ref{sub:Smearing}).

\begin{figure}
\begin{centering}
\includegraphics[width=0.15\textwidth]{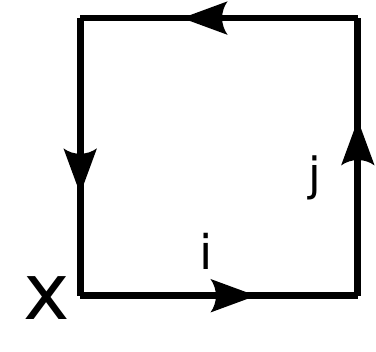}\qquad{}\qquad{}\qquad{}\includegraphics[width=0.2\textwidth]{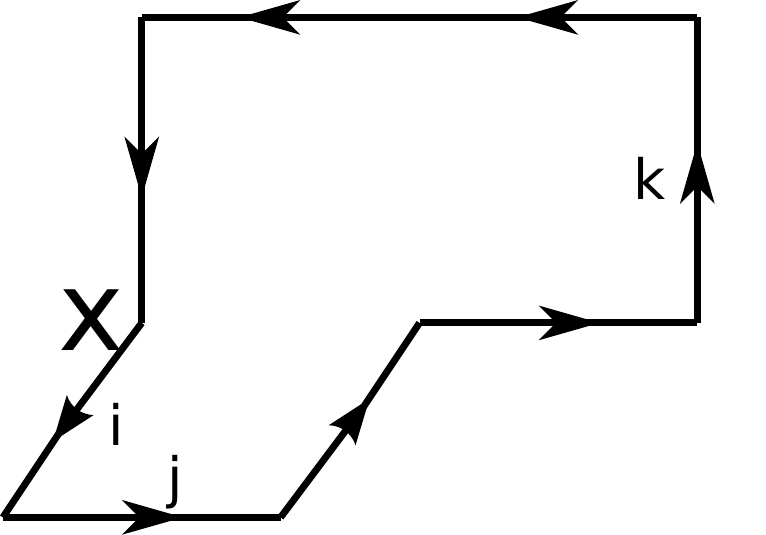}\protect\caption{Plaquette $P_{ij}(x)$ and gauge-loop $\mbox{\ensuremath{\mathcal{C}}}_{ijk}(x)$
used in the interpolating field of $0^{-+}$-glueball.}
\label{fig: gauge-loops} 
\par\end{centering}

\end{figure}

The basic interpolating fields for glueballs are built from
gauge link loops that respect the parity quantum number of the respective
state. For the $0^{++}$-glueballs we use a sum of gauge plaquettes
\begin{equation}
O_{\text{gb}^{++}}(x)=\text{Tr}\left[P_{12}(x)+P_{23}(x)+P_{31}(x)\right],
\end{equation}
where $P_{ij}$ denotes a plaquette in the $i$-$j$ direction (see Fig. \ref{fig: gauge-loops}).
For the pseudo-scalar glueball $0^{-+}$ we use 
\begin{equation}
O_{\text{gb}^{-+}}(x)=\sum_{R\in\mathbf{O}_{h}}[\text{Tr}\left[\mathcal{C}(x)\right]\text{-}\text{Tr}\left[P\mathcal{C}(x)\right],
\end{equation}
where the sum is over all rotations of the cubic group $\mathbf{O}_h$ and $P\mathcal{C}$ is the parity conjugate of loop $\mathcal{C}$; 
  it is depicted in Fig. \ref{fig: gauge-loops}.

The basic interpolating fields for the mesons are
\begin{align}
O_{\text{a-f}_{0}}(x)=&\bar{\lambda}(x)\lambda(x), \qquad 
O_{\text{a-}\eta'}(x)=\bar{\lambda}(x)\gamma_{5}\lambda(x).
\end{align}
When inserted into the correlation matrix, Wick contractions of these fields lead to connected and disconnected pieces
\begin{equation}
 \langle\bar{\lambda}(x)\Gamma\lambda(x)\bar{\lambda}(y)\Gamma\lambda(y)\rangle=
 \text{Tr}\left[\Gamma\Delta(x,x)\right]\text{Tr}\left[\Gamma\Delta(y,y)\right]-2\text{Tr}\left[\Gamma\Delta(x,y)\Gamma\Delta(y,x)\right],
\end{equation}
where $\Delta(x,y)$ denotes the propagator from $x$ to $y$ (spin and group indices suppressed) and $\Gamma$ represents $\mathbb{1}$ or $\gamma_5$.

The interpolating field of the gluino-glue state is given by 
\begin{equation}
 O_{\text{g-g}}^{\alpha}(x)=\sum_{i<j=1}^3\sigma_{ij}^{\alpha\beta}\text{Tr}\left[P_{ij}(x)\lambda^{\beta}(x)\right],
\end{equation}
with $\sigma_{\mu\nu}= \frac{\text{i}}{2}[\gamma_\mu,\gamma_\nu].$ 

\subsection{Smearing Methods} \label{sub:Smearing}

In order to improve the signal-to-noise ratio and to obtain different
interpolating fields $O_i$ for the variational method, we use APE-smearing \cite{Albanese:1987ds} for the gauge links $U_\mu(x)$ and gauge
invariant Jacobi smearing \cite{Allton:1993wc,Best:1997qp} for the gluino field $\lambda(x)$.

An APE-smearing step is given by 
\begin{equation} 
U'_{\mu}(x) =U_{\mu}(x)+\epsilon_{\text{APE}}\sum_{\nu=\pm1,\nu\neq\mu}^{\pm3}U_{\nu}^{\dagger}(x+\hat\mu)U_{\mu}(x+\hat{\nu})U_{\nu}(x),
\end{equation}
where $\epsilon_{\text{APE}}$ is a parameter of the smearing method that needs to be optimized to maximize the signal-to-noise ratio.
The APE-smearing step can be iteratively applied to obtain higher smearing levels.
The gluino field is smeared by applying the Jacobi smearing function:
\begin{equation}
 \lambda'^{\beta b}(\vec{x},t)=\sum_{\vec{y}}F_{\beta b,\alpha a}(\vec{x},\vec{y},t)\lambda^{\alpha a}(\vec{y},t),
\end{equation}
where the smearing function $F_{\beta b,\alpha a}(\vec{x},\vec{y},t)$ is given by
\begin{equation}
F_{\beta b,\alpha a}(\vec{x},\vec{y},t)=\delta_{\vec{x},\vec{y}}\,\delta_{\beta\alpha}+
\delta_{\beta\alpha}\sum_{i=1}^{N_{\text{J}}}\left(\kappa_{\text{J}}\sum_{\mu=1}^{3}\left[\delta_{\vec{y},\vec{x}+\hat\mu}U_{\mu}(\vec{x},t)+
\delta_{\vec{y}+\hat{\mu}}U_{\mu}^{\dagger}(\vec{x},t)\right]\right)^{i}_{ba}, \label{eq: Jacobi smearing function}
\end{equation}
with the smearing parameter $\kappa_{\text{J}}$  and the smearing level $N_{\text{J}}$.

\subsection{Identifying the eigenvalues} \label{sub:identification}
The errors of the generalized eigenvalues that are used to fit the masses and
the final errors of the masses are estimated by resampling the data using the Jackknife method.

\begin{figure}
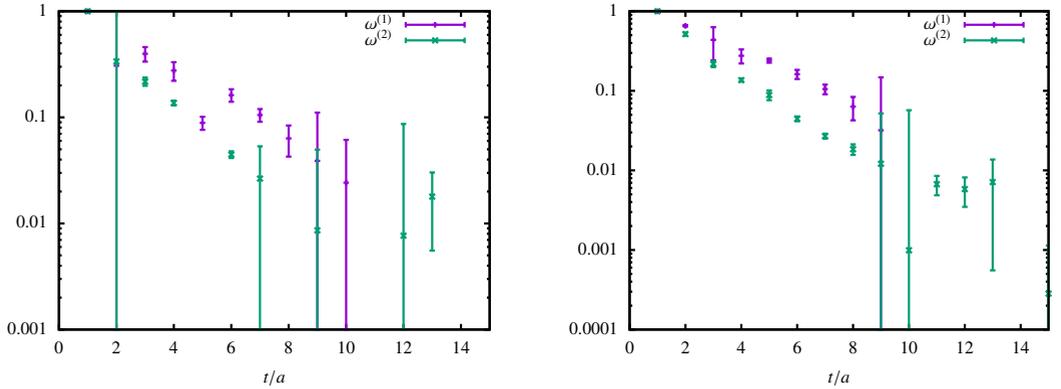


\subfigure{\resizebox{0.43\paperwidth}{!}{\input{allEigenvalues_reduced.tex}}}
\subfigure{\resizebox{0.43\paperwidth}{!}{\input{allEigenvalues_reduced_ordered.tex}}}
\caption{Left: Generalized eigenvalues in the $0^{++}$-channel at $\beta=1.9$, $\kappa=0.1433$
sorted by magnitude, right: sorted by scalar products of eigenvectors. Six interpolating fields were used in the GEVP; for better visibility only two of the six eigenvalues are displayed. 
Note that the eigenvalues at $t=5$ have been misidentified when sorted by magnitude. The identification using scalar products allows to use the data up to $t=8$.}
\label{eq:sorted_eigenvalues}

\end{figure}

Fitting an exponential to the $n$-th generalized eigenvalue $\omega^{(n)}(t,t_0)$ requires identifying it out of all the generalized eigenvalues at $t$ on each Jackknife sample.
Originally it was proposed to identify the eigenvalues at each $t$ by their magnitude, $\omega^{(0)}>\omega^{(1)}>\omega^{(2)},\ldots$ .
However, due to limited statistics the eigenvalues fluctuate on the Jackknife samples. 
If the fluctuations of two or more eigenvalues overlap, it is not suitable any longer to identify them by their magnitude. 
In this situation we use a different approach, which is to identify the generalized eigenvalues by their corresponding generalized eigenvectors $\vec{v}^{\ (n)}$  in the following way:
\begin{enumerate}
\item Solve the GEVP for all $t$, find a $t_1$ where the  fluctuations of the $\omega^{(n)}(t_1,t_0)$ are small enough such that they do not overlap.
$t_1$ should be chosen as large as possible, so that the eigenvectors have sufficiently stabilized. Sort the eigenvalues by their magnitude as originally proposed.
The corresponding eigenvectors are called $\vec{w}^{(n)}$.

\item Solve the GEVP at all other $t$, call the eigenvectors $\vec{v}^{\ (n)}(t,t_0)$.

\item For each $t$ calculate  a matrix of scalar products $M_{mn}=|\vec{w}^{(m)}\cdot\vec{v}^{\ (n)}(t,t_0)|$.
\item
\begin{enumerate}
\item Scan for largest entry $M_{n_\text{max}m_\text{max}}$ in $M$.
\item Identify $\omega^{(m_\text{max})}(t,t_0)$ to be the eigenvalue corresponding to $\vec{v}^{(n_{\text{max}})}(t,t_0)$.
\item Delete $m_\text{max}$-th row and $n_\text{max}$-th column from $M$.
\item Repeat 3. until all eigenvalues are identified.
\end{enumerate}
\end{enumerate}
A comparison of this method with the earlier one is shown in  Fig. \ref{eq:sorted_eigenvalues}.

\subsection{Optimization of the smearing parameters}

\begin{figure}
\begin{centering}
\sidecaption
\resizebox{0.45\paperwidth}{!}{\input{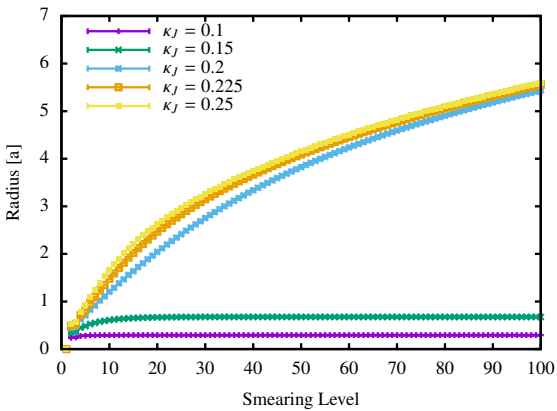}}
\caption{Measurement of the Jacobi smearing radius for different values of
$\kappa_{\text{J}}$ at $\beta=1.9,$ and $\kappa=0.14415$. 
For $\kappa_{\text{J}}\geq0.2$ the smearing function diverges for $N_\text{J}\rightarrow\infty$. For $\kappa_{\text{J}}\leq0.15$ it converges.}
\label{fig: Smearing Radius}


\end{centering}

\end{figure}

In order to determine the optimal values for the Jacobi smearing parameters $\kappa_{\text{J}}$ and $N_\text{J}$, we measured
the smearing radius defined as 
\[
R_{\text{J}}^{2}=\dfrac{\sum_{\vec{x}}\left|\vec{x}\right|^2\left|F(\vec{x},0)\right|^2}{\sum_{\vec{x}}\left|F(\vec{x},0)\right|^2}.
\]
The measurement shows that there is a critical parameter $0.15<\kappa_{\text{J}}^{\text{c}}<0.2$ (see Fig. \ref{fig: Smearing Radius}). For
values above $\kappa_{\text{J}}^{\text{c}}$ the smearing function
$(\ref{eq: Jacobi smearing function})$ diverges in the limit $N_\text{J}\rightarrow \infty$, while for smaller
values it converges. Since we use Jacobi smearing to create new interpolating
fields from the basic ones, it is more suitable to choose a value for
$\kappa$ in the diverging regime. In order to keep numerical errors
small, we chose $\kappa_{\text{J}}^{\text{c}}=0.2$ which is just above
the critical $\kappa_{\text{J}}^{\text{c}}$. Optimizing the signal-to-noise
ratio, we chose to use the smearing levels 0, 40 and 80.
In principle more smearing levels could be used, but from our experience it makes the identification of the eigenvalues (see Sect. \ref{sub:identification}) unstable and therefore does not improve the results.

A similar analysis was also done for the APE-smearing. In the gluino-glue measurements we chose $\epsilon_{\text{APE}}=0.4$ and the smearing levels $0,5,15,25,..,95$.
In the glueball measurements we chose the smearing levels 10, 50 and 80.

\subsection{Exploiting mixing}
Operators with the same quantum numbers are expected to mix. 
We expect the $\text{a-f}_{0}$ meson to mix with the $0^{++}$-glueball since they share the same quantum numbers. 
The same is true for the a-$\eta'$ meson and the $0^{-+}$-glueball. 
This mixing allows both, mesonic and gluonic, interpolators to be used in the correlation matrix (\ref{eq:Correlation-Matrix})
for the extraction of a mixed state. The full correlation matrix then has the following form:

\begin{equation}
C(t)=\left(\begin{array}{cc}
\left\langle O_{\text{gb}^{++}}(t)O^\dagger_{\text{gb}^{++}}(0)\right\rangle  & \left\langle O_{\text{gb}^{++}}(t)O^\dagger_{a-f_{0}}(0)\right\rangle \\
\left\langle O_{a-f_{0}}(t)O^\dagger_{\text{gb}^{++}}(0)\right\rangle  & \left\langle O_{a-f_{0}}(t)O^\dagger_{a-f_{0}}(0)\right\rangle 
\end{array}\right).
\end{equation}
Here each entry is a submatrix consisting of the different interpolating fields for each operator. 
Since the mesonic and the gluonic operators are very different from one another, the mutual overlap is expected to be small and 
using this larger correlation matrix should improve the signal drastically.

\section{Preliminary Results}
We have used lattices for $\mbox{\ensuremath{\beta=1.9}}$ at four different hopping parameters $\mbox{\ensuremath{\kappa\in\{0.1433,0.14387,0.14415,0.14435\}}}$. 
For each value of $\kappa$ there are more than $10.000$ lattice gauge configurations available of which only every $8$th configurations has been measured so far. 
Our results are therefore only preliminary. 

\begin{figure}[!thb] 
\begin{centering}

\sidecaption
\resizebox{0.55\paperwidth}{!}{\input{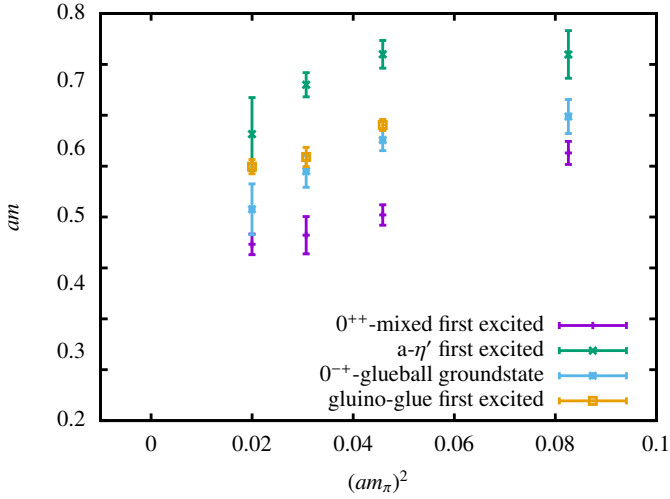}}
\caption{Preliminary results for $\beta=1.9$. The extracted masses are plotted against the measured squared mass of the unphysical adjoint pion $m_\pi$. 
The gluino mass scales proportionally to $m_\pi^2$, thus the chiral limit is at $m_\pi^2=0$ \cite{Munster:2014cja}. }
\label{fig:chiralextrapolation}
\end{centering}
\end{figure}

The results (see Fig. \ref{fig:chiralextrapolation}) indicate that the variational 
method in combination with different smearing levels and different interpolating fields can indeed be used to extract the masses of the excited states. 
As expected, we observe a much cleaner signal in the $0^{++}$-channel when using both, mesonic and gluonic, interpolators than when using them separately.
Curiously, in the $0^{-+}$-channel the correlation matrix is block-diagonal within errors, i.e. there seems to be no mixing between the a-$\eta'$ meson and the $0^{-+}$-glueball. 
The groundstate mass of the $0^{-+}$-glueball seems to be in the range of the excited states.
These two findings might be an indication that the $0^{-+}$-component of the lowest chiral multiplet only consists of a-$\eta'$ and that
the $0^{-+}$-glueball actually belongs to a different, heavier multiplet as was already proposed in Ref. \cite{Bergner:2015adz}.

For the final results the statistics will be increased and a detailed analysis of autocorrelations will be done. A chiral extrapolation to the critical $\kappa_\text{c}$ will be performed.
Previous results \cite{Bergner:2015adz} indicate that the ensemble at $\beta=1.9$ is already close to the continuum limit.
To provide an estimate of lattice discretization errors, we will also analyze a coarser lattice spacing at $\beta=1.75$.

\section*{Acknowledgements}

The authors gratefully acknowledge the Gauss Centre for Supercomputing
(GCS) for providing computing time for a GCS Large-Scale Project on
the GCS share of the supercomputer JUQUEEN at Jülich
Supercomputing Centre (JSC). GCS is the alliance of the three national
supercomputing centres HLRS (Universit\"at
Stuttgart), JSC (Forschungszentrum Jülich),
and LRZ (Bayerische Akademie der Wissenschaften), funded by the German
Federal Ministry of Education and Research (BMBF) and the German State
Ministries for Research of Baden-Württemberg
(MWK), Bayern (StMWFK) and Nordrhein-Westfalen (MIWF). Further computing
time has been provided on the supercomputers JURECA at JSC and on the compute cluster PALMA and NWZPHI of the University of Münster. 
This work is supported by the Deutsche Forschungsgemeinschaft (DFG)
through the Research Training Group ``GRK 2149: Strong and Weak Interactions - from Hadrons to Dark Matter''.

\bibliography{Bibliography_spire}

\begin{thebibliography}{12}

\bibitem{Curci:1986sm}
G.~Curci, G.~Veneziano, Nucl. Phys. \textbf{B292}, 555 (1987)

\bibitem{Bergner:2016sbv}
G.~Bergner, S.~Catterall, Int. J. Mod. Phys. \textbf{A31}, 1643005 (2016),
  \texttt{arXiv:1603.04478}

\bibitem{Veneziano:1982ah}
G.~Veneziano, S.~Yankielowicz, Phys. Lett. \textbf{113B}, 231 (1982)

\bibitem{Farrar:1997fn}
G.~R. Farrar, G.~Gabadadze, M.~Schwetz, Phys. Rev. \textbf{D58}, 015009 (1998),
  \texttt{arXiv:hep-th/9711166}

\bibitem{Bergner:2015adz}
G.~Bergner, P.~Giudice, I.~Montvay, G.~Münster, S.~Piemonte, JHEP \textbf{03},
  080 (2016), \texttt{arXiv:1512.07014}

\bibitem{thiscontrib102}
S.~Ali, G.~Bergner, H.~Gerber, P.~Giudice, I.~Montvay, G.~Münster,
  S.~Piemonte, P.~Scior, \emph{{Ward identities in N = 1 supersymmetric SU(3)
  Yang-Mills theory on the lattice}}, in \emph{Proceedings,
  \href{http://inspirehep.net/record/1425631}{35th International Symposium on
  Lattice Field Theory (Lattice2017)}: Granada, Spain}, to appear in EPJ Web
  Conf., \texttt{arXiv:1710.01000}

\bibitem{Morningstar:2003gk}
C.~Morningstar, M.~J. Peardon, Phys. Rev. \textbf{D69}, 054501 (2004),
  \texttt{arXiv:hep-lat/0311018}

\bibitem{Luscher:1990ck}
M.~Lüscher, U.~Wolff, Nucl. Phys. \textbf{B339}, 222 (1990)

\bibitem{Albanese:1987ds}
M.~Albanese et~al. (APE Collaboration), Phys. Lett. \textbf{B192}, 163 (1987)

\bibitem{Allton:1993wc}
C.~R. Allton et~al. (UKQCD Collaboration), Phys. Rev. \textbf{D47}, 5128
  (1993), \texttt{arXiv:hep-lat/9303009}

\bibitem{Best:1997qp}
C.~Best et~al., Phys. Rev. \textbf{D56}, 2743 (1997),
  \texttt{arXiv:hep-lat/9703014}

\bibitem{Munster:2014cja}
G.~Münster, H.~Stüwe, JHEP \textbf{05}, 034 (2014), \texttt{arXiv:1402.6616}

\end{thebibliography}


\end{document}